\documentclass[twocolumn,reprint,aps,prd,nofootinbib,floatfix]{revtex4-1}
\usepackage[utf8]{inputenc}
\usepackage[utf8]{inputenc}
\usepackage{amsmath}
\usepackage{amsfonts}
\usepackage{amssymb}
\usepackage[left=2cm,right=2cm,top=2cm,bottom=2cm]{geometry}
\usepackage{braket}
\usepackage{dsfont}
\usepackage{hyperref}

\usepackage{graphicx}
\usepackage{tikz}
\usetikzlibrary{arrows,decorations.pathmorphing,backgrounds,positioning,fit,petri,shapes}
\usepackage{lipsum}

\usepackage{multirow}
\begin{document}

\title{Non-parametric Cosmology with Cosmic Shear}

\begin{abstract}
We present a method to measure the growth of structure and the background geometry of the Universe -- with no \emph{a priori} assumption about the underlying cosmological model. Using Canada-France-Hawaii Lensing Survey (CFHTLenS) shear data we simultaneously reconstruct the lensing amplitude, the linear intrinsic alignment amplitude, the redshift evolving matter power spectrum, $P \left(k,z \right)$, and the co-moving distance, $r \left(z \right)$. We find that lensing predominately constrains a single global power spectrum amplitude and several co-moving distance bins. Our approach can localise precise scales and redshifts where Lambda-Cold Dark Matter (LCDM) fails -- if any. We find that below $z = 0.4$, the measured co-moving distance $r \left(z \right)$ is higher than that expected from the Planck LCDM cosmology by $\sim 1.5 \sigma $, while at higher redshifts, our reconstruction is fully consistent. To validate our reconstruction, we compare LCDM parameter constraints from the standard cosmic shear likelihood analysis to those found by fitting to the non-parametric information and we find good agreement.
\end{abstract}

\author{Peter L.~Taylor}
\email{peterllewelyntaylor@gmail.com}
\author{Thomas D.~Kitching}
\author{Jason D.~McEwen}
\affiliation{Mullard Space Science Laboratory, University College London, Holmbury St.~Mary, Dorking, Surrey RH5 6NT, UK}
\date{16 November 2018}
\maketitle

\section{Introduction}
\par The leading cosmological model, Lambda-Cold Dark Matter (LCDM), is purely phenomenological. There is no widely accepted physical mechanism that explains the existence of dark matter, nor the accelerated expansion of the Universe. For this reason, in addition to measuring the LCDM parameters to ever great precision, we must test -- rather than assume -- the LCDM paradigm. 
\par To achieve this aim, we take advantage of an effect called gravitational lensing. As light from distant galaxies travels to Earth its path is distorted by the gravitational pull of intervening mass. This causes small changes in the observed ellipticities and sizes of the galaxies. The coherent signal this induces, which is only detectable by measuring the shape of many galaxies, is called cosmic shear. 
\par From the first detections of cosmic shear in 2000-2001~\cite{rhodes2001detection, bacon2000detection, wittman2000detection}, studies have entered the realm of precision cosmology \cite{heymans2013cfhtlens, troxel2017dark, kitching20143d, hildebrandt2017kids, hikage2018cosmology}. With the advent of Stage IV lensing experiments including Euclid\footnote{\url{http://euclid-ec.org}} \cite{laureijs2010euclid}, WFIRST\footnote{\url{https://www.nasa.gov/wfirst}} \cite{spergel2015wide} and LSST\footnote{\url{https://www.lsst.org}} \cite{anthony4836large}, constraints on cosmological parameters will shrink by a further order of magnitude \cite{refregier2010euclid}.
\par Every cosmic shear study to date has assumed a cosmological model, before inferring the values of the model's parameters. We follow an alternative approach and reconstruct the expansion history of the Universe and the evolution of large scale structure formation independently from any cosmological model. This is in a similar spirit to~\cite{tegmark2002separating}.
\par There are many ways to extract cosmological information from the shear catalog. By far the most popular technique is the Gaussian likelihood analysis of the shear two-point statistic \cite{heymans2013cfhtlens, troxel2017dark, kitching20143d, hildebrandt2017kids, hikage2018cosmology}. Other statistics include: peak counts~\cite{peel2017cosmological, jain2000statistics}, higher order statistics~\cite{semboloni2010weak, fu2014cfhtlens} and Minkowski functionals~\cite{petri2013cosmology}. 
\par We choose to use the shear two-point statistic. Even though this only accesses the Gaussian information of the shear field, a large body of work exists to provide rigorous requirements on systematics to ensure unbiased results \cite{massey2012origins, PLTpreapring, amara2007optimal}. 
\par Using data from the Canada-France-Hawaii Lensing Survey (CFHTLenS) we reconstruct the lensing amplitude $\mathcal{A}_G$, the linear intrinsic alignments amplitude, $\mathcal{A}_{IA}$, the co-moving distance, $r \left(z \right)$\footnote{Formally we measure the co-moving angular distances but for a spatially-flat universe this is equivalent to the co-moving distance.} and the matter power spectrum, $P \left(k,z \right)$. We refer to this as non-parametric cosmology because this information can always be measured without ever needing to assume a cosmological model parametrized in terms of a small number of physical parameters. Meanwhile we refer to the information contained in these amplitudes and functions as the non-parametric information. This study directly builds off \cite{PLTpreapring} where we found the precise scales and redshifts where cosmic shear is sensitive to the power spectrum and co-moving distance. 
\par Furthermore once we have extracted the non-parametric information it is possible to use this to test \emph{any} cosmological model, without having to repeat the lensing analysis itself. To verify the fidelity of our non-parametric reconstruction we infer the LCDM parameters directly from the non-parametric information and compare to the standard cosmic shear likelihood analysis. 
\par While cosmic shear extracts both distance and structure growth information, measurements of Baryonic Acoustic Oscillations (BAO)~\cite{alam2017clustering} and Type Ia supernovae (SNe Ia)~\cite{riess20162} already constrain the cosmic distance to within a few percent at low redshifts. A disagreement in the inferred expansion history between our non-parametric cosmic shear reconstruction these two other two measurements would indicate the presence of systematics in one or more of the experiments. 
\par
\begin{figure*}
 \includegraphics[width=0.79\linewidth]{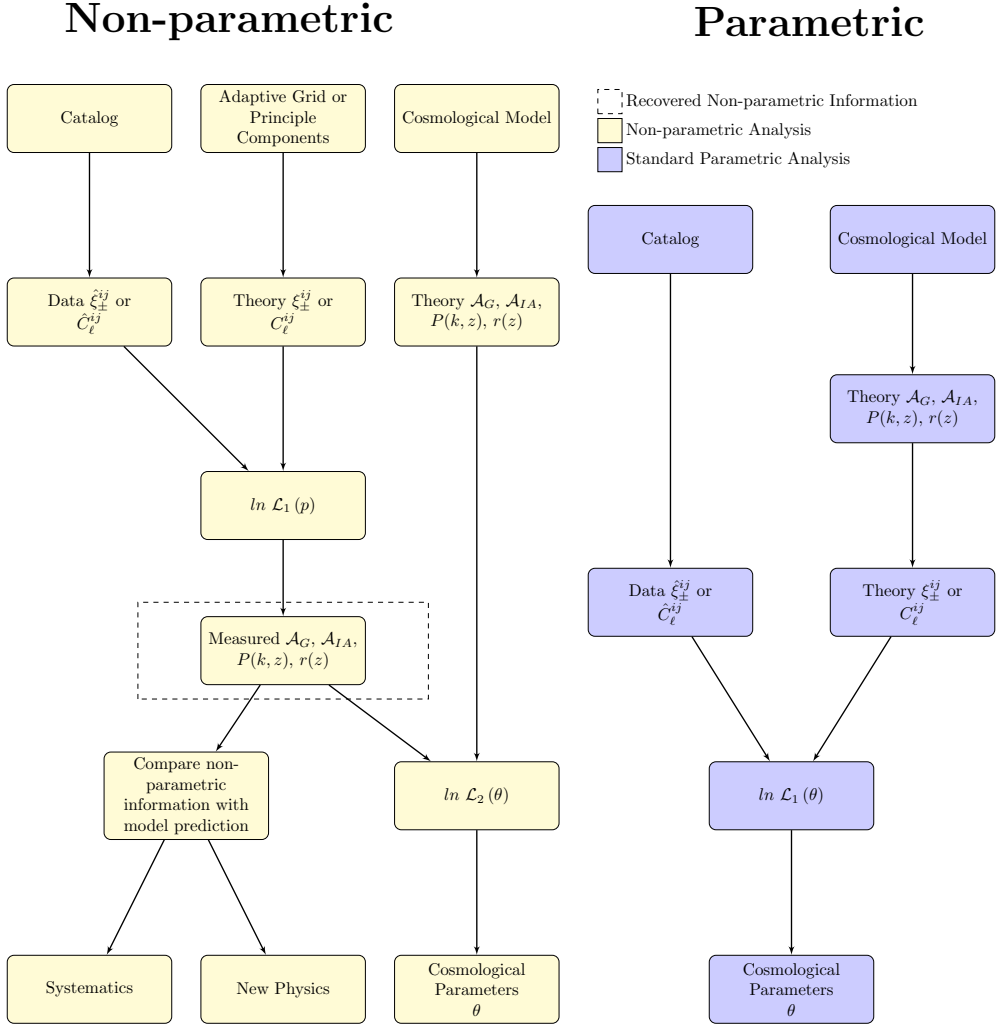}
\caption{The main steps of the non-parametric reconstruction and the standard parametric inference. These techniques are discussed in detail in Sections~\ref{sec:Lensing Formalism}-\ref{sec:PI}. The non-parametric reconstruction has a number of desirable features. 1. We recover the expansion and structure growth history, with no need to assume a cosmological model. 2. Once the non-parametric information is recovered, we can test any cosmological model without needing to re-compute lensing observables. 3. Comparing the non-parametric and parametric reconstructions pinpoints the precise redshifts and scales where the cosmological model fails -- if any. Knowing this could help narrow the search for previously unidentified systematics (see the second paragraph of Section~\ref{sec:future prospects}). After a thorough search -- if the discrepancies are believed to be physical -- this would indicate precisely how the Universe deviates from LCDM in a fully non-parametric way.}
\label{fig:flowchart}
\end{figure*}

 In Section~\ref{sec:Formalism and Data} we review the cosmic shear formalism, discuss the CFHTLenS data and present our technique for extracting the non-parametric information. A flowchart outlining the main steps of the non-parametric reconstruction is shown in Figure~\ref{fig:flowchart}. The results are presented in Section~\ref{sec:results} and the future prospects of our technique are discussed in Section~\ref{sec:future prospects}.

\section{Formalism and Data} \label{sec:Formalism and Data}
\subsection{Cosmic Shear Formalism} \label{sec:Lensing Formalism}

Cosmic shear extracts cosmological information from the correlation in the ellipticity between pairs of galaxies.  The ellipticity is written as a complex number so that $\epsilon = \epsilon_1 + i \epsilon_2$. Then for pairs of galaxies the tangential and cross-ellipticity $\epsilon_{+, \times}$ are the tangential and cross ellipticities in the frame joining a pair of galaxies $\{ a, b\}$. To extract more information, we also bin galaxies radially in tomographic redshift bins. Then the correlation function for angular separation $\theta$ is given by:

\begin{equation}\label{eqn:xidata}
\hat{\xi}_{\pm}^{ij}(\theta) = \frac{\sum w_a w_b \left[ \epsilon_{+}^i (x_a) \epsilon_{+}^j (x_b) \, \pm \, \epsilon_\times^i (x_a) \epsilon_\times^j (x_b)
\right]}{
\sum w_a w_b },
\end{equation}
where $w_a$ and $w_b$ are weights and $i$ and $j$ denote the tomographic bin number, and the sums are over all galaxy pairs. 
\par The correlation function is related to the convergence power spectrum, $C_{GG}^{ij}(\ell)$, by:

\begin{equation} \label{eqn:xiGG}
\xi_{\pm, {\rm GG}}^{ij}(\theta) = \frac{1}{2\pi}\int d\ell \,\ell \,C_{GG}^{ij}(\ell) \, J_{\pm}(\ell \theta), 
\end{equation}
where $J_{+}(\ell \theta)$ is the zeroth order Bessel function of the first kind and $J_{-}(\ell \theta)$ is the fourth order Bessel function of the first kind.
\par Taking the Limber, flat-sky and flat-Universe assumptions, the convergence power spectrum can be written as:

\begin{equation} \label{eqn:Cl}
C_{GG}^{ij}(\ell) = \int_0^{\infty} dk \, \frac{\ell ^ 3}{k ^ 4} \, 
\frac{w_i(\ell / k )w_j(\ell / k )}{a \left( \ell / k \right)^2} \, P \left( k, \frac{\ell}{k} \right), 
\end{equation}
where $a$ is the expansion factor, $P$ is the matter power spectrum and the lensing efficiency in harmonic space, $w_i$, is
\begin{equation} 
w_i(\ell / k ) = \frac{3 H_0^2 \Omega_{\rm m}}{2c^2} \, \int_0^{\infty}\, dr[z']\ n_i(r[z']) \,
F \left(r[z'],\ell / k \right) , 
\end{equation}
where 
\begin{equation} \label{eqn:lenskern}
F\left( r, r'\right) = \frac{r-r'}{rr'}
\end{equation}
is the lensing kernel and $H_0$, $\Omega_m$, $c$, $a$ and $n_i$ respectively denote the present day Hubble parameter, the  fraction energy density of matter compared to the critical density, the speed of light in vacuum, the expansion factor and the radially distribution of observed galaxies in tomographic bin $i$.
\par Equation~\ref{eqn:Cl} - \ref{eqn:lenskern} are not the the usual expression for the power spectrum, lensing efficiency and lensing kernel given in~\cite{bridle2007dark,heymans2013cfhtlens}, but it is easy to derive by making the change of variable $r=\ell/k$ (see the Appendix of~\cite{kitchinglimits} for more details). We make this transformation because during the non-parametric reconstruction we do not enforce monotonicity of $r(z)$, so it is no longer an appropriate way to label the cosmic time at which the matter power spectrum is evaluated.
\par We have developed our own code to compute the convergence spectrum in this notation and verified against {\tt GLaSS}~\cite{PLTtesting} and the shear code integrated into the modular cosmology package {\tt Cosmosis}~\cite{cosmosis}. Throughout this work we use {\tt CAMB}~\cite{camb} to generate the linear power spectrum and expansion history and {\tt Halofit}~\cite{halofit} to generate the non-linear power spectrum. All correlation functions are computed using {\tt Nicea}~\cite{kilbinger2009dark}. All modules are integrated with {\tt Cosmosis} and called through this interface.  
%

\subsection{Intrinsic Alignments} \label{sec:Intrinsic Alignments}
In addition to instrumental systematics, astrophysical systematics must also be accounted for. The dominant contribution comes from intrinsic alignments caused by the tidal alignment of galaxies around large dark matter halos. This dampens the lensing spectrum and leads to two additional terms in the theoretical expression for the correlation function. An `II term' accounts for the intrinsic tidal alignment of galaxies around massive dark matter halos. Meanwhile the `GI term' accounts for the anti-correlation between tidally aligned galaxies at low redshifts and weakly lensed galaxies at high redshift. Then the theoretical correlation function is a sum of the lensing and intrinsic alignment auto and cross-correlation functions, given by:
\begin{equation} \label{eqn:xiGI}
{\xi}_{\pm}^{ij}(\theta) = \xi_{\pm,{\rm II}}^{ij}(\theta) +\xi_{\pm,{\rm GI}}^{ij}(\theta) +\xi_{\pm,{\rm GG}}^{ij}(\theta).
\end{equation}
We follow the linear intrinsic alignment model originally given in~\cite{hirata2004intrinsic} and used in~\cite{heymans2013cfhtlens}. In this model the theoretical expression for II and GI correlation functions are:
\begin{equation}\label{eqn:xitheory}
\xi_{\pm,{\rm II/GI}}^{ij}(\theta) = \frac{1}{2\pi}\int d\ell \,\ell \,C_{\rm II/GI}^{ij}(\ell) \, J_{\pm}(\ell \theta) , 
\end{equation}
where the II spectrum, $C_{\rm II}^{ij}(\ell)$, is given by:
\begin{equation}
C_{\rm II}^{ij}(\ell) = \int_0^{r_{\rm H}} dr \, 
\frac{n_i(r)n_j(r)}{r^2} \, P_{\rm II} \left( \frac{\ell}{r},r \right),
\label{eqn:CII} 
\end{equation}
where the II matter power spectrum is:
\begin{equation} \label{eqn:PII}
P_{\rm II}(k,z) =  F^2(z) P(k,z)
\end{equation}
and
\begin{equation} \label{eqn:Fz} 
F(z) = - A_I C_1 \rho_{\rm crit} \frac{\Omega_{\rm m}}{D(z)}.
\end{equation}
Here $\rho_{\rm crit}$ is the critical density of the Universe, $D(z)$ is the growth factor and $C_1 = 5 \times 10^{-14} h^{-2} M_\odot^{-1} {\rm Mpc}^3$ is chosen so that the fiducial value of the intrinsic alignment amplitude, $A_I$, is unity~\cite{brown2002measurement}. 
Meanwhile the GI matter power spectrum is:
\begin{equation}
C_{\rm GI}^{ij}(\ell) = \int_0^{r_{\rm H}} dr \, 
\frac{q_i(r)n_j(r) + n_i(r)q_j(r) }{r^2} \, P_{\rm GI} \left( \frac{\ell}{r},r \right),
\label{eqn:CGI} 
\end{equation}
where the lensing efficiency kernel, $q_i$ is defined as:
\begin{equation}
q_i(r) = \frac{3 H_0^2 \Omega_{\rm m}}{2c^2} \frac{r}{a(r)}\int_r^{r_{\rm H}}\, dr'\ n_i(r') 
\frac{r'-r}{r'}, 
\label{eqn:qk} 
\end{equation}
and the GI spectrum is:
\begin{equation} \label{eqn:PGI}
 P_{\rm GI}(k,z) =  F(z) P (k,z).
\end{equation}
\par Formally we should have made the transformation $r=\ell/k$ to label the matter power spectra in equations~\ref{eqn:PII} and \ref{eqn:PGI} in terms of $k$ as we did for the shear spectra defined in equation~\ref{eqn:Cl}. However, because the contamination from intrinsic alignments is so small relative to the statistical error for CFHTLenS, we ignore this complication for the time being and just use the co-moving distance of the fiducial cosmology in these expressions for the remainder of this work. We have also chosen not to include an `IG' term to account for the correlation between foreground shear with background intrinsic alignments. This is non-zero due to photometric redshift error. Nevertheless the magnitude of the IG term is usually an order of magnitude smaller than the II term~\cite{brown2011polarization}, so it can be safely ignored at this stage.

\subsection{CFHTLenS Data} \label{sec:CFHTLens Data}
We use public shear data from the Canada-France-Hawaii Lensing Survey, CFHTLenS. This is a lensing survey covering $154 \text{ deg} ^2$ in five optical bands. Galaxies were observed at a median redshift $z_m = 0.70$ with an effective weighted number density of $n_{\rm eff} =11$ galaxies per square arcmin. Catalogs were produced by combining weak lensing data processing from {\tt THELI}~\cite{erben2013cfhtlens}, shear measurements from {\tt Lensfit}~\cite{miller2013bayesian} and photometric redshift estimates from PSF-matched photometry~\cite{hildebrandt2012cfhtlens}.
\par We use the same $6$ tomographic bins as ~\cite{heymans2013cfhtlens} for galaxies in the redshift range $0.2 < z < 1.30$. Bins were defined by dividing galaxies into the redshift ranges: $z_1 \in [0.2, 0.39]$, $z_2 \in [0.39, 0.58]$, $z_3 \in [0.58, 0.72]$, $z_4 \in [0.72, 0.86]$, $z_5 \in [0.86, 1.02]$, and $z_6 \in [1.02, 1.30]$. Bins were smoothed by Gaussian kernel with dispersion $\sigma_z =0.04 \left( 1 + z \right)$ to account for the photometric redshift uncertainty. We use the same angular bins as~\cite{heymans2013cfhtlens}.

\subsection{Likelihood and Covariance Matrix}\label{sec:Likelihood and Covariance Matrix}
To extract the cosmological information from the shear catalog, we assume a Gaussian likelihood:
\begin{equation} \label{eq:gauss}
\text{ln } \mathcal{L}_1\left(p \right) = - \frac{1}{2} \sum_{a,b} \left[ D_a - T_a\left( p \right) \right] C_{ab}^{-1} \left[ D_b - T_b \left( p\right) \right],
\end{equation}
where $ D_a$ is the data vector composed of the observed $\hat{\xi}_{\pm}^{ij}$ and $T_a\left( p \right)$ is formed from the theoretical prediction of $\xi_{\pm}^{ij}$ given parameters $p$ and $C_{ab}^{-1}$ is the inverse of the covariance matrix. 
\par The data and theory vectors are taken to be the correlation functions defined in equations \ref{eqn:xidata} and \ref{eqn:xitheory} respectively. In the standard cosmic shear likelihood analysis, the parameters $\{ p \}$ are taken to be the cosmological model parameters and a set of nuisance parameters (e.g. the amplitude of the intrinsic alignments $A_I$). In our non-parametric analysis we will take the parameters $\{ p \}$ to be a set of amplitudes that encode information about the lensing amplitude, the intrinsic alignment amplitude, the co-moving distance and the power spectrum (see Sections~\ref{sec:AG} and \ref{sec:PC} for more details).
\par Meanwhile we use the publicly available covariance matrix from the CFHTLenS survey\footnote{This covariance matrix is available for download from \url{http://www.cfhtlens.org/astronomers/cosmological-data-products}} integrated into {\tt Cosmosis 2pt} module (see~\cite{heymans2013cfhtlens} for more details). The matrix is generated from the N-body lensing simulations of~\cite{harnois2012gravitational}. Since the covariance is generated from noisy realizations we apply the Anderson-Hartlap correction when inverting the covariance~\cite{hartlap2007your, anderson1958introduction}.
\par We use the Markov Chain Monte Carlo (MCMC) sampler {\tt emcee}~\cite{foreman2013emcee} to sample the likelihood and perform the inference.

\subsection{Non-parametric Information Extraction}\label{sec:NP}
The cosmic shear spectrum defined in equation~\ref{eqn:Cl} is only sensitive to the cosmology of the Universe through  the power spectrum, $P(k,z)$, the co-moving distance, $r(z)$ and a set of lensing amplitudes. If we assume the linear intrinsic alignment model, then the lensing amplitudes are: an overall shear amplitude, $\mathcal{A}_G$, and an intrinsic alignment amplitude, $\mathcal{A}_{IA}$. From equations~\ref{eqn:Cl} and ~\ref{eqn:Fz} we see $A_G \propto \Omega_m H_0^2$ and $\mathcal{A}_{IA} \propto A_I \Omega_m$.
\par The main idea of this paper is to divide the power spectrum and co-moving distance into cells/bins and simultaneously measure the amplitudes of these cells and the lensing amplitudes. In detail, we generate a fiducial\footnote{We have found that all the results presented in this paper are insensitive to changing the fiducial cosmology parameters by up to $15 \%$} power spectrum, co-moving distance and lensing amplitudes using the CFHTLenS best fit cosmology from~\cite{heymans2013cfhtlens}. We divide the power spectrum into a set of logarithmically spaced cells in the ($k$,$z$) plane $\{ P_i \}$ and the co-moving distance into a set of cells $\{r_i\}$. Perturbing the lensing amplitudes, the power spectrum cells and co-moving distance cells, we form a vector of amplitudes $p = \left(\mathcal{A} (\mathcal{A}_{G\rm }), \mathcal{A} (\mathcal{A}_{\rm IA}),\{ \mathcal{A} (P_i) \}, \{\mathcal{A} (r_j) \} \right) $ where the amplitudes are defined relative to the fiducial cosmology, so that:
\begin{equation}
\begin{aligned}
\mathcal{A} (\mathcal{A}_{\rm G}) &= \mathcal{A}_{\rm G} / \mathcal{A}_{G}^{\rm fid} \\
\mathcal{A} (\mathcal{A}_{\rm IA}) &= \mathcal{A}_{\rm IA} / \mathcal{A}_{IA}^{\rm fid} \\
\mathcal{A} (P_i) &= P_i / P^{\rm fid}_i \\
\mathcal{A} (r_j) &= r_j / r^{\rm fid}_j
\end{aligned}
\end{equation}
where $P_i$ is the power spectrum in cell $i$, $P^{\rm fid}_i$ is the fiducial power spectrum inside cell $i$, $r_j$ is the fiducial co-moving distance in cell $j$, and $r^{\rm fid}_j$ is the fiducial co-moving inside cell $j$. New GG, GI and II spectra are defined as functions of the amplitude vector, p, and written as $C_{\rm GG}^{ij}(\ell, p )$, $C_{\rm GI}^{ij}(\ell, p)$ and $C_{\rm II}^{ij}(\ell, p)$. Substituting the perturbed spectra into equations~\ref{eqn:xiGG} and~\ref{eqn:xiGI} forms the theory vector in the Gaussian likelihood defined in equation~\ref{eq:gauss} and we will infer the posterior distribution on the amplitude vector, $p$, using CFHTLenS shear data.
\par To start, we divide the power spectrum into $100$ cells on a $10 \times 10$ grid in $k$ and $z$. The co-moving distance is divided into $10$ cells in $z$. With the lensing and intrinsic alignment amplitude, this leaves us with $112$ amplitudes to measure. To perform the inference we first compress the amplitude vector, $p$, down to a more manageable size using two different data compression regimes, which are discussed in the next two sections. 

\subsection{Adaptive Grid Compression}\label{sec:AG}
\begin{figure}
\centering
\includegraphics[width = 90mm]{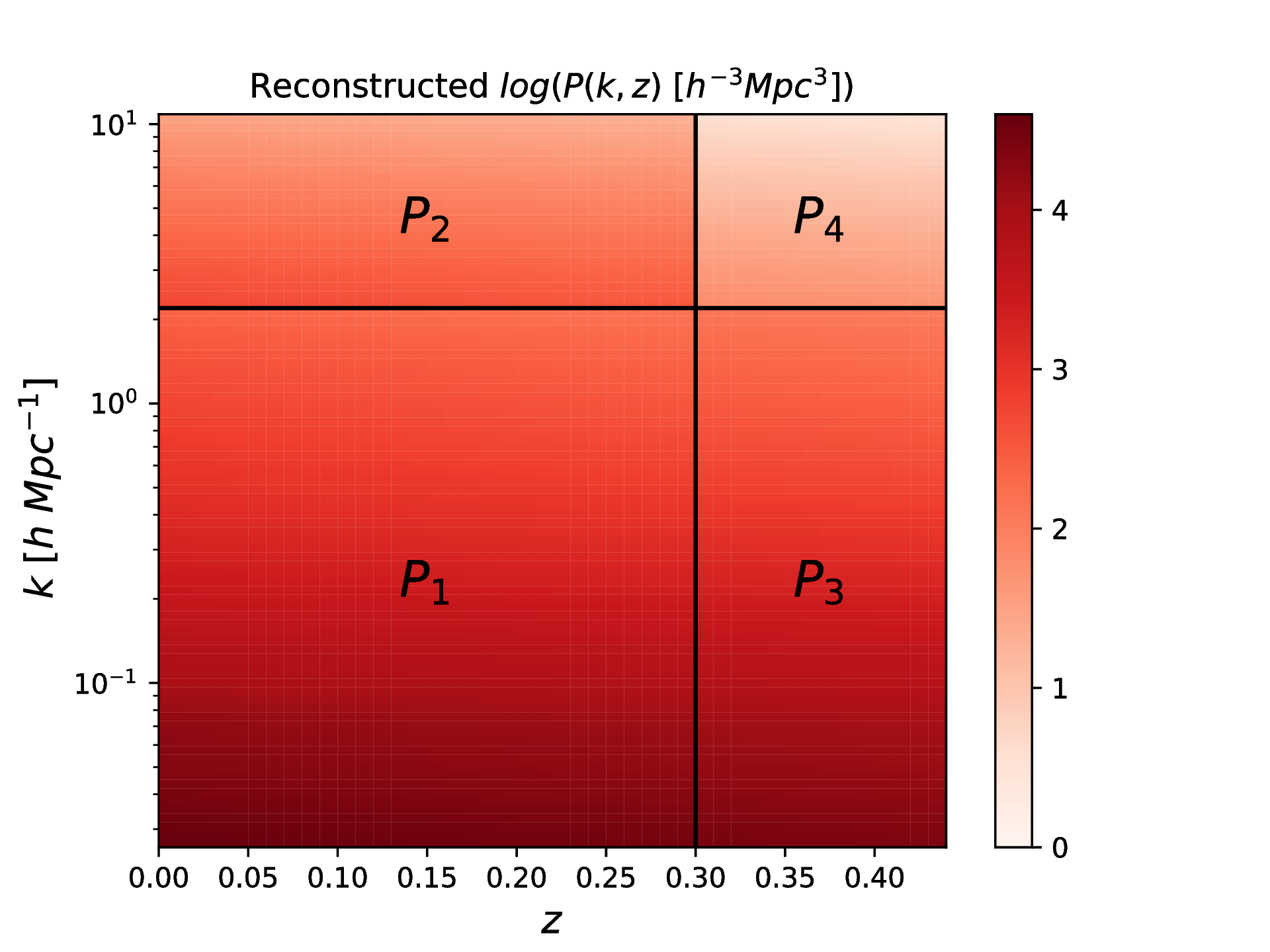}
\includegraphics[width = 90mm]{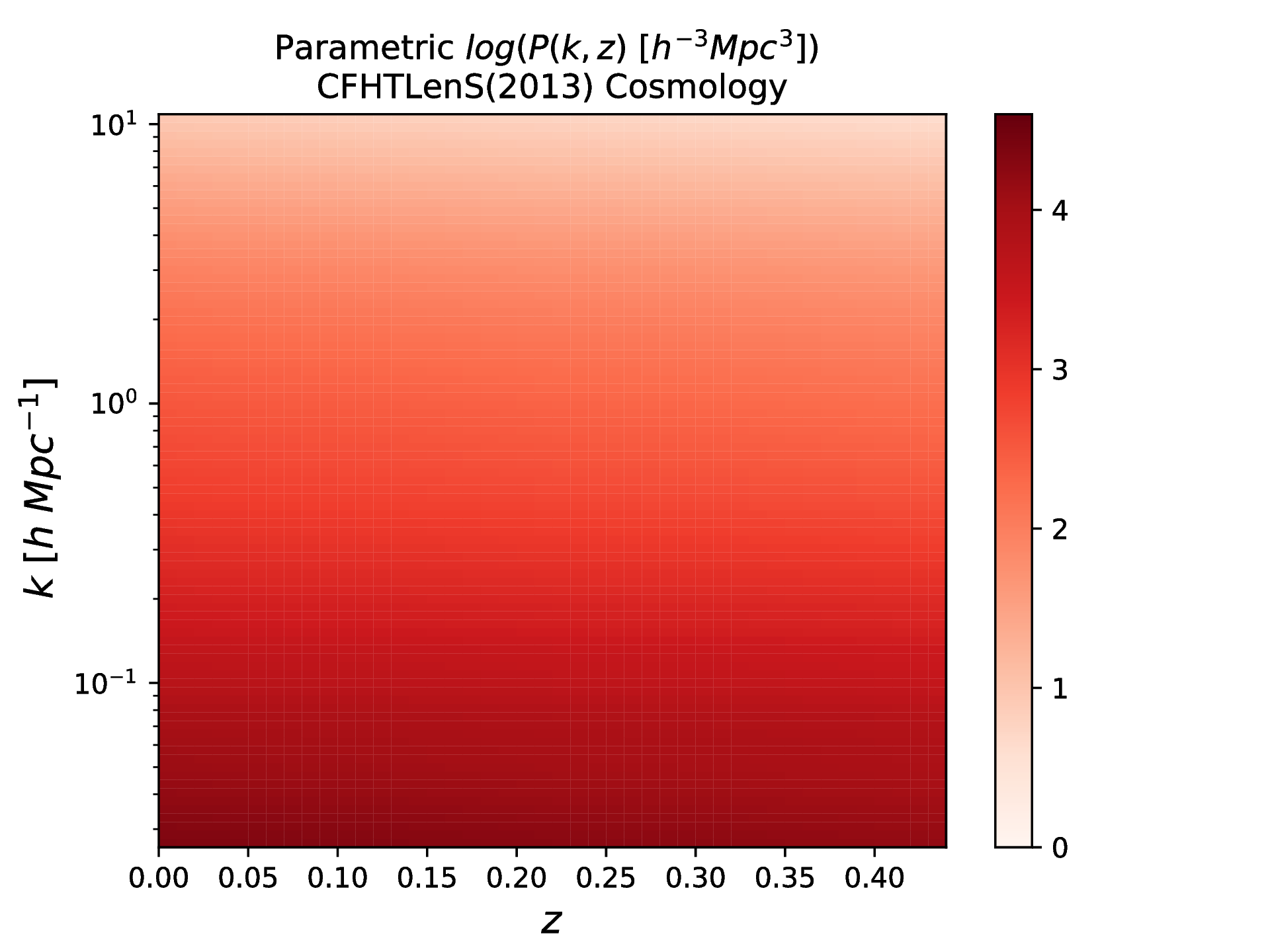}
\caption{ {\bf Top:} Best fit non-parametric reconstructed power spectrum. This is the first non-parametric reconstruction of the time evolving matter power spectrum from shear data. Currently the errors on this reconstruction are very large (see Figure~\ref{fig:amps_grid}), but these will shrink by a factor of $\sim 20-25$ with a Stage-IV experiment. {\bf Bottom:} Power spectrum generated by {\tt CAMB}~\cite{camb} and {\tt HALOFIT}~\cite{halofit} using CFHTLenS~\cite{heymans2013cfhtlens} best fit LCDM parameters. The non-parametric and parametric reconstruction are in agreement.}
\label{fig:p_kf}
\end{figure}

\begin{figure}
\centering
\includegraphics[width = 90mm]{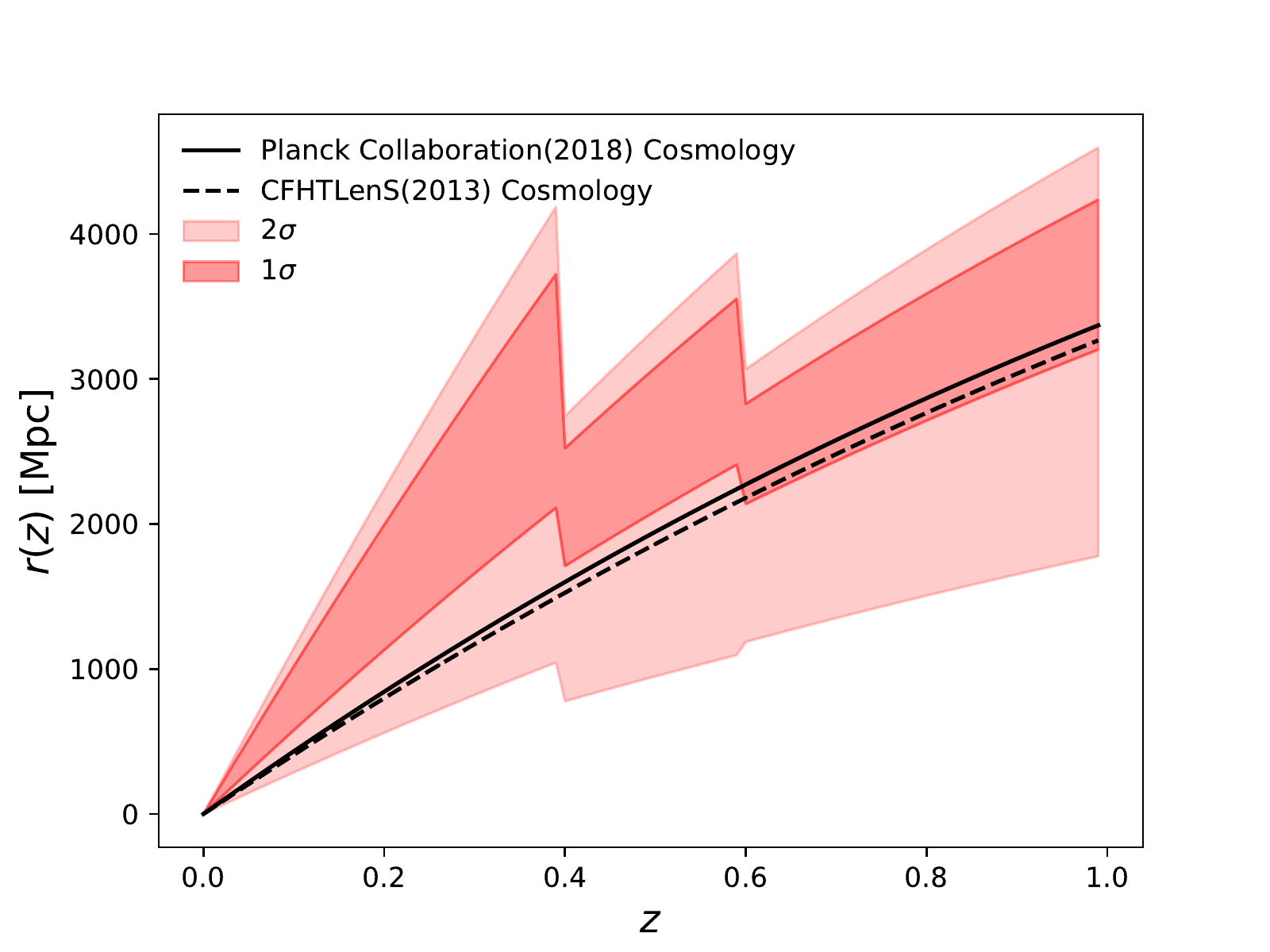}
\caption{The $1 \sigma$ and $2 \sigma$ constraints on the reconstructed non-parametric $r(z)$. The jumps in the constraints are due to binning. Unlike the power spectrum, this is fairly well constrained with CFHTLenS data. We also plot the parametric $r(z)$ using a LCDM cosmology with the best fit parameters from CFHTLenS~\cite{heymans2013cfhtlens} and the Planck 2018 combined analysis (including BAO)~\cite{planck18}. In the CFHTLenS (2013) study the Hubble parameter, $h_0$, is given a tight prior (see Section~\ref{sec:cosmo agrid}). Repeating the CFHTLenS (2013) analysis with a flat prior on $h_0$, we find it is only constrained in the range $\left( 0.4, 1.2 \right)$ at the $1 \sigma$ level. Since $r(z)$ is proportional to $h_0 ^{-1}$, there is no internal tensions between our non-parametric distance measurement and the parametric analysis. Nevertheless, below $z = 0.4$ the non-parametric reconstruction is in mild $\sim 1.5 \sigma$ tension with the the Planck combined cosmology $r(z)$.}
\label{fig:r_z}
\end{figure}

We employ the Fisher matrix formalism to assess cosmic shear's sensitivity to each amplitude. The Fisher matrix is given by:
\begin{equation} \label{eqn:error}
F_{ij} = \sum_{a,d} \frac{\partial D_a}{\partial p_i} C_{ab}^{-1} \frac{\partial D_b}{\partial p_j}, 
\end{equation}
where $D_a$ is the data vector formed from the correlation functions $\hat{\xi}_{\pm} \left(\theta \right)$,  $C_{ab}^{-1}$ is the covariance matrix given in equation~\ref{eq:gauss}, while $p_i$ and $p_j$ are amplitudes in the vector $p$. It is now convenient to define the information content contained in a set of amplitudes $\{ p_i \}$ as:
\begin{equation}
I = \sum_{i \in \{ p_i\}} 1/ F_{ii}.
\end{equation}
\par In our adaptive grid regime, we combine adjacent power spectrum and co-moving distance cells so that we are left with a much smaller set of cells that each contain roughly the same amount of information, $I$. 
\par Specifically, the power spectrum is divided into four cells that each contain roughly $25\%$ of the remaining information. The cell boundaries are plotted in Figure~\ref{fig:p_kf}. Meanwhile, the co-moving distance is divided into three cells in $z$, shown in Figure~\ref{fig:r_z}, so that each contain roughly a third of the co-moving distance information. Altogether our compressed amplitude vector, $p$, is formed of two lensing amplitudes, four power spectrum cells and three co-moving cells:
\begin{equation}
p = \left(\mathcal{A} (\mathcal{A}_{G\rm }), \mathcal{A} (\mathcal{A}_{\rm IA}),\{ \mathcal{A} (P_i) \}, \{\mathcal{A} (r_j) \} \right)
\end{equation}
where $i \in [1,4]$ and $j \in [1,3]$.
\par Since this paper is a proof of concept, this compression is in no way optimal and several arbitrary choices (e.g. the number of power spectrum cells) have been made. Since optimizing this compression is survey specific, we leave optimizing this procedure to a future work with new data.

\subsection{Principle Component Compression}\label{sec:PC}

We also use the popular principle component analysis (PCA) compression technique, as an alternative to the adaptive grid. Given a high-dimensional data set, PCA compression works by finding an orthogonal set of vectors which contain the majority of the variance in the data and only saving this information. In our case we want to solve the opposite problem and minimize the variance by finding the linear combinations of amplitudes $p_i \in p$ which are the most tightly constrained by the shear data.  
\par The predicted variance is encoded in the Fisher matrix, $F$. Specifically the covariance between amplitudes is $C_p = F^{-1}$. Rotating into the eigenbasis yields:
\begin{equation} \label{eqn:PCArot}
C_p = P^T D P,
\end{equation}
where $D$ is a diagonal matrix of eigenvalues, $\{ \lambda_i\}$, and $P$ is a rotation matrix with columns formed from the associated eigenvectors. 
\par Arranging the eigenvalues in ascending order, the corresponding eigenvectors are called the principle components (PCs). We take the first $10$ PCs to form our compressed amplitude vector: 
\begin{equation}
p = (\{ PC_i\})
\end{equation}
where $i \in [1,10]$.
\par If the lensing likelihood was exactly Gaussian (see~\cite{sellentin2017insufficiency, sellentin2018skewed} for a discussion of why it is not) then these components would contain $77\%$ of the total inverse variance. Again, we stress that this is an arbitrary choice which will need fine-tuning in future studies.

\subsection{Cosmological Parameter Inference from Non-parametric Information}\label{sec:PI}
Normally cosmological parameters, $\theta$, are found straight from the shear data by sampling from $\text{ln } \mathcal{L}_1 (\theta)$, defined in equation~\ref{eq:gauss}. We now discuss a technique to extract the cosmological parameters directly from the measured non-parametric amplitudes. This is used to validate the non-parametric reconstruction. In the future this technique can be used to test a large number of cosmological models quickly and consistently \emph{without repeating the lensing analysis}.

\par Using the MCMC chains from the non-parametric reconstruction as data, we form the Gaussian likelihood
\begin{equation} \label{eqn:likenonparam}
\text{ln } \mathcal{L}_2\left(\theta \right) = - \frac{1}{2} \sum_{a,b} \left[ \hat p_a - T_a\left( \theta \right) \right] \hat C_{ab}^{-1} \left[ \hat p_b - T_b \left( \theta  \right) \right],
\end{equation}
where $\hat p$ is the mean of amplitude vector over all samples in the chain, and the covariance, $\hat C$, between amplitudes is given by
\begin{equation}
\hat C_{ab} = \langle \left( p_a - \hat p_a \right) \left( p_b - \hat p_b \right) \rangle,
\end{equation}
where the average is taken over all samples in the chain.
\par The theory vector, $T_a \left( \theta  \right)$, depends on which compression regime was used. In the adaptive grid case the theoretical lensing amplitudes are
\begin{equation}
\begin{aligned} \label{eqn:Atheory}
\mathcal{A}^{\rm Th}\left( \mathcal{A}_{\rm G} \right) &= \frac{\Omega_m H_0 ^ 2 \sigma_8}{\Omega_m^{\rm fid} H_0^{\rm fid\text{ }2} \sigma_8^{\rm fid}} \\
\mathcal{A}^{\rm Th}\left( \mathcal{A}_{\rm IA} \right) &= \frac{\Omega_m A_I \sigma_8}{\Omega_m^{\rm fid} A_I^{\rm fid} \sigma_8^{\rm fid}}.
\end{aligned}
\end{equation}
We have pulled out an overall scaling amplitude of the power spectrum, $\sigma_8$, so the theoretical power spectrum amplitude inside cell $i$ must be appropriately rescaled\footnote{We have found that if we do not do pull out an overall scaling factor, we do not accurately recover the tails of the posterior on $\theta$. We are free to make this choice provided we rescale the power spectrum amplitudes appropriately.}. It is given by:
\begin{equation} \label{eqn:ptheory}
\mathcal{A}^{\rm Th}\left( P_i \right) = \left( \frac{\sigma_8^{\rm fid}}{\sigma_8 } \right) ^2 \langle P_i \left( \theta \right) / P_i ^{\rm fid} \rangle,
\end{equation}
where the average is taken over all sampled points in the cell. The theoretical co-moving distance amplitude inside cell $j$ is:
\begin{equation} \label{eqn:rtheory}
\mathcal{A}^{\rm Th}\left( r_j \right) = \langle r_i \left( \theta \right) / r_j ^{\rm fid} \rangle,
\end{equation}
and the average is again over all points in the cell. In summary, the theory and data vectors for the adaptive grid technique are given by:
\begin{equation}
\begin{aligned}
T =& \left(\mathcal{A}^{\rm Th}\left( \mathcal{A}_{\rm G} \right), \mathcal{A}^{\rm Th}\left( \mathcal{A}_{\rm IA} \right), \{ \mathcal{A}^{\rm Th}\left( P_i \right) \}, \{ \mathcal{A}^{\rm Th}\left( r_j \right) \} \right) \\
\hat p =& \left( \langle \mathcal{A}\left( \mathcal{A}_{\rm G} \rangle \right), \langle \mathcal{A}\left( \mathcal{A}_{\rm IA} \right) \rangle, \{ \langle \mathcal{A}\left( P_i \right) \rangle \}, \{ \langle \mathcal{A}\left( r_j \right) \rangle \} \right),
\end{aligned}
\end{equation}
where $i \in [1,4]$ and $j \in [1,3]$ and the averages are taken over all samples in the reconstructed amplitude chain which is found from sampling from $\text{ln } \mathcal{L}_1 (p)$. 
\par In the PCA compression case, we take the theoretical amplitudes defined in equations~\ref{eqn:Atheory} - \ref{eqn:rtheory}, and rotate these into PCA space using the rotation matrix $P$, defined in~\ref{eqn:PCArot}. Explicitly we compute
\begin{equation} \label{eq:rotation}
\mathcal{A}^{{\rm PC},{\rm Th}} = I + R \left(\mathcal{A}^{\rm Th} - I \right)
\end{equation}
The first $10$ rows of the rotation matrix, $R$, are the same as $P$ while the remaining rows are set to zero since we are assuming no contribution from the remaining PCs. The theoretical amplitude vector, $\mathcal{A}^{\rm Th}$, appearing in equation~\ref{eq:rotation} is defined in terms of the amplitudes written in equations~\ref{eqn:Atheory}-\ref{eqn:rtheory} and is given by:
\begin{equation}
\mathcal{A}^{\rm Th} = \left(\mathcal{A}^{\rm Th}\left( \mathcal{A}_{\rm G} \right), \mathcal{A}^{\rm Th}\left( \mathcal{A}_{\rm IA} \right), \{ \mathcal{A}^{\rm Th}\left( P_i \right) \}, \{ \mathcal{A}^{\rm Th}\left( r_j \right) \} \right)
\end{equation}
where $i \in [1,100]$ and $j \in [1,10]$ run over the original cells. Finally, $I$ is a dimension $112$ vector with all entries equal to unity. We subtracted this before rotation in equation~\ref{eq:rotation} because the PCA amplitudes are defined relative to unity.
In summary, the theory and data vectors for the PCA technique are given by:
\begin{equation}
\begin{aligned}
T =& \left( \{ \mathcal{A}^{\rm PC,Th}_i \} \right) \\
\hat p =& \left( \{ \langle  \mathcal{A}_i  \rangle \} \right)
\end{aligned}
\end{equation}
for $i \in [1,10]$ and the average, just like in the adaptive grid case, is over all samples in the chain.
\par Using both compression techniques, we can now sample from likelihood defined in~\ref{eqn:likenonparam} to compute the posterior distribution on the cosmological parameters $\theta$. The process is schematically shown in Figure~\ref{fig:flowchart}.

\section{Results} \label{sec:results}

\subsection{Adaptive Grid Reconstruction}
\begin{figure*}
\includegraphics[width=1.\linewidth]{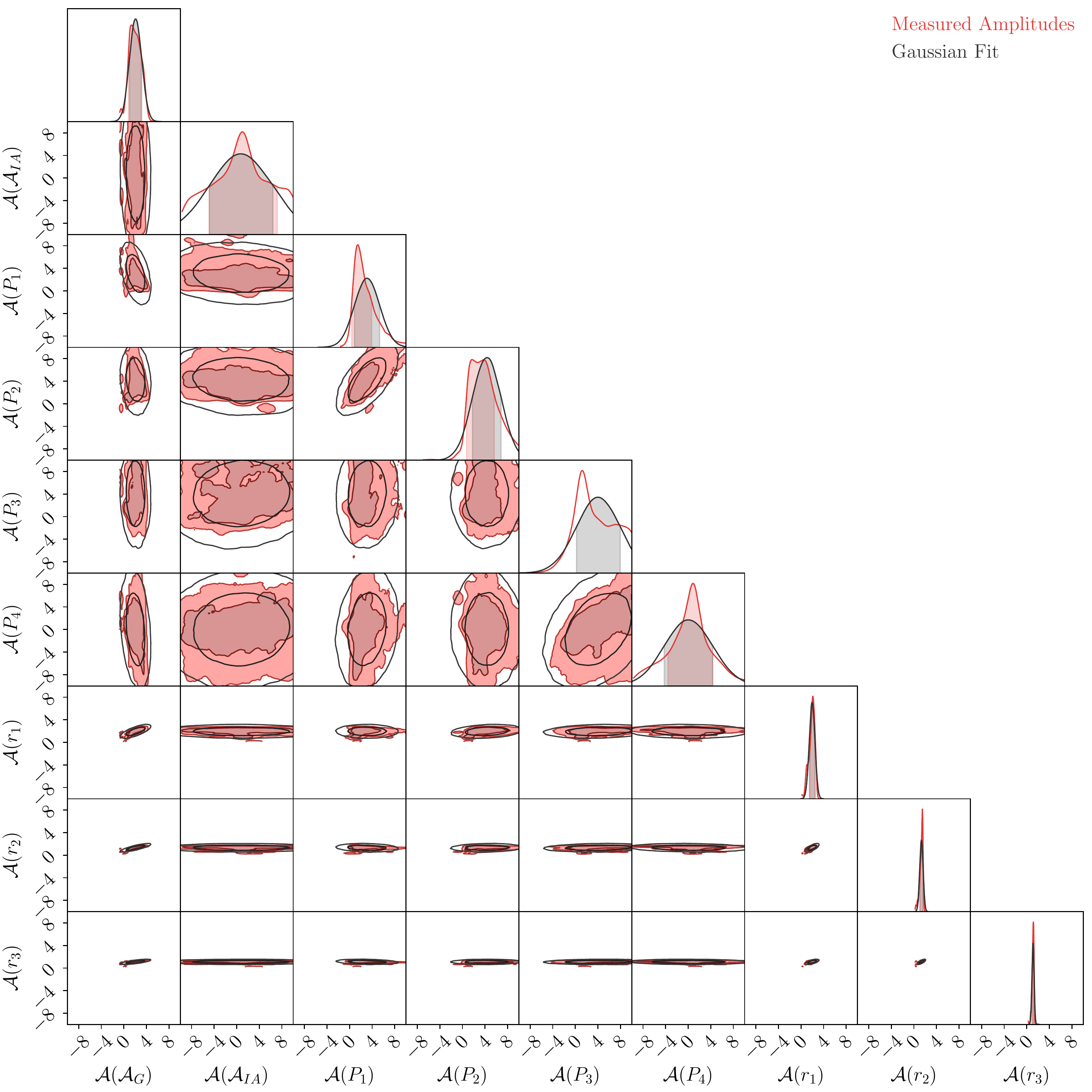}
\caption{The reconstructed amplitudes using the adaptive grid technique. We measure the amplitude of the lensing signal $\mathcal{A}_{\rm G}$, the intrinsic alignment amplitude $\mathcal{A}_{\rm IA}$, the amplitude of four power spectrum bins in $k$ and $z$ and the amplitude of three co-moving distances bins in $z$. The bin boundaries for the power spectrum and co-moving distance are illustrated in Figures~\ref{fig:p_kf} and ~\ref{fig:r_z}. Only the amplitude of the lensing signal and the co-moving distance are well constrained. There is a degeneracy between the lensing amplitude and the co-moving distance amplitudes because both are strongly dependent on $\Omega_m$ and $h_0$. We plot the Gaussian distribution that we have fit to the chains to form the likelihood $\text{ln } \mathcal{L}_1 (p)$, in gray. All corner plots in this work are produced using {\tt ChainConsumer}~\cite{chainconsumer} using the default kernel density estimate settings.}
\label{fig:amps_grid}
\end{figure*}

Sampling from the likelihood $\text{ln } \mathcal{L}_1 (p)$, we measure two lensing amplitudes, four power spectrum amplitudes and three co-moving distance amplitudes. The recovered posterior distribution of the amplitudes is plotted in Figure~\ref{fig:amps_grid}. Only the lensing amplitude and the co-moving distance amplitudes are tightly constrained. The amplitude of individual matter power spectrum cells are hardly constrained at all.  
\par Figure~\ref{fig:p_kf} shows the best fit non-parametric power spectrum. For comparison, a parametric power spectrum generated from the best fit CFHTLenS LCDM cosmological parameters~\cite{heymans2013cfhtlens} is shown. The two are in good agreement, particularly since the error bars on the non-parametric reconstruction are so large (see Figure~\ref{fig:amps_grid}).
\par In Figure~\ref{fig:r_z} we plot the non-parametric reconstruction of the co-moving distance $r(z)$. The LCDM prediction generated with the best fit parameters from both CFHTLenS~\cite{heymans2013cfhtlens} and the Planck 2018 combined analysis (including BAO)~\cite{planck18} are shown for comparison. While it may appears that there is an internal inconsistency between our non-parametric reconstruction and the parametric analysis of CFHTLenS data in~\cite{heymans2013cfhtlens}, this is just due to the choice of prior on $h_0$ in the analysis presented in~\cite{heymans2013cfhtlens} (see the discussion in the caption of Figure~\ref{fig:r_z}). 
\par For $z < 0.4$, our non-parametric reconstruction of $r(z)$ is in $~\sim 1.5 \sigma$ tension with the Planck LCDM predictions. In the range $0.4 <z <0.6$, this drops to a $~\sim 1 \sigma$ tension, while for $z>0.6$ there is no tension at all. 
\par The discrepancy between the non-parametric $r(z)$ and the Planck LCDM reconstruction is unlikely to be caused by poor photometric redshift error estimation because we would expect these to get worse at higher redshifts, not lower redshifts where the tension occurs. We leave a thorough non-parametric study of systematic effects to future work.
\par It is also pointed out in~\cite{unified}, that positive values for the intrinsic alignment parameter, $A_I$, are favored by CFHTLenS. This is the opposite sign to what is expected by theory and could point to lingering systematic effects in the shear catalog. Whatever the cause of the co-moving distance tension, we intend to investigate this further with data from other surveys.

\subsection{PCA Reconstruction}

\begin{figure*}
\includegraphics[width=1.\linewidth]{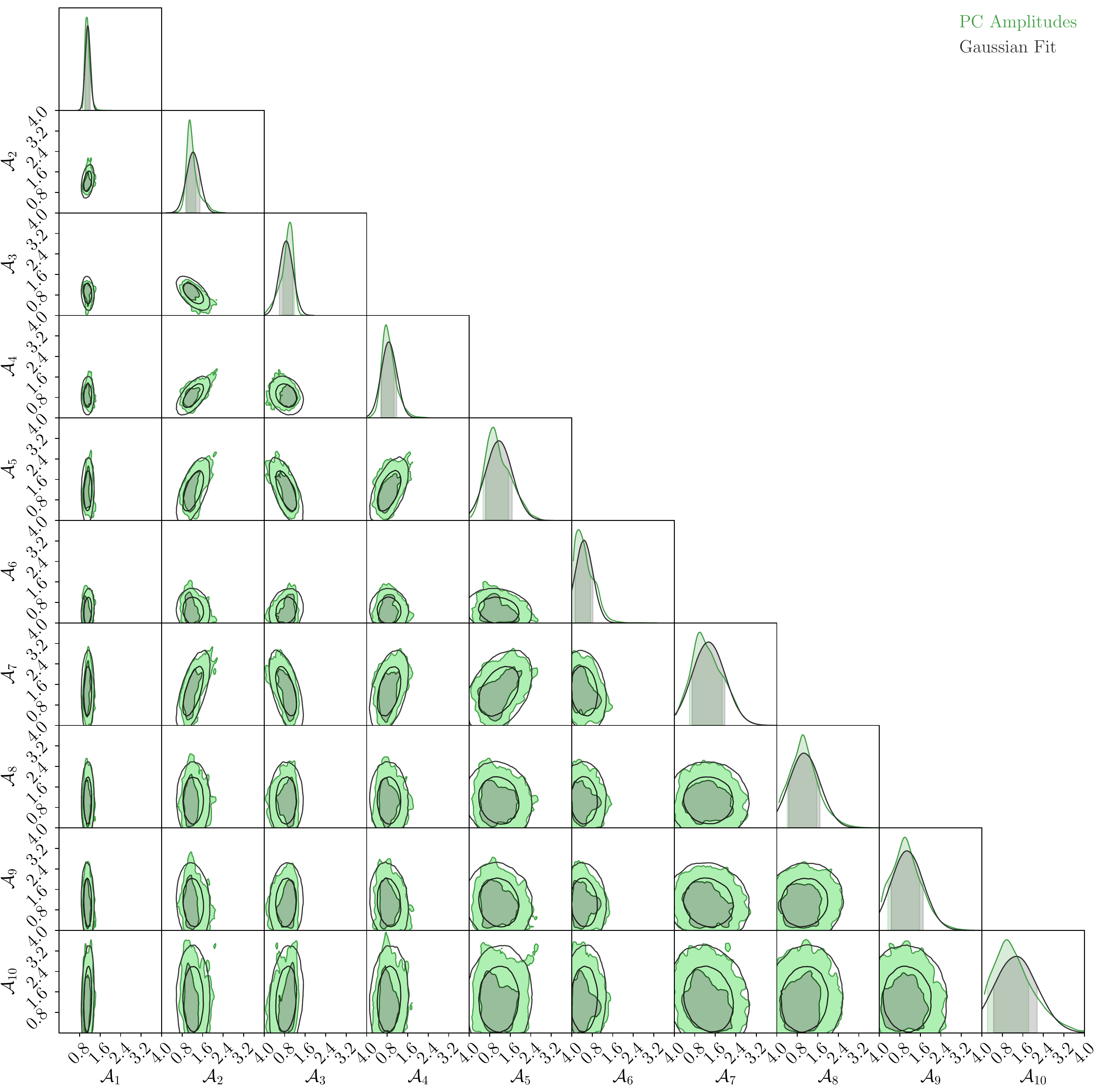}
\caption{The posterior distribution on the measured PC amplitudes. The first two PCs are as expected from the Fisher matrix prediction, but because the likelihood is non-Gaussian, the constraints on the other PCs are up to twice as wide as expected. We plot the Gaussian distribution that we have fit to the chains to form the likelihood $\text{ln } \mathcal{L}_1 (p)$, in gray.}
\label{fig:amps_pca}
\end{figure*}

Sampling from the likelihood $\text{ln } \mathcal{L}_1 (p)$, we measure the first $10$ PC amplitudes. The posterior distribution on the amplitudes is shown in Figure~\ref{fig:amps_pca}. 
\par Since the lensing likelihood is non-Gaussian~\cite{sellentin2018skewed}, the posteriors do not agree with the Fisher expectation, particularly past the first two PCs. The constraints on the first PC amplitude is only $10 \%$ wider than expected, but the constraints on the third PC amplitude are twice as wide as expected. Degeneracies between a few PC amplitudes (e.g between $\mathcal{A}_1$ and $\mathcal{A}_2$) are also present. By construction these are absent from the Fisher prediction.
\par Although it is possible to rotate the PC amplitudes back into the ($k$, $z$) plane to reproduce Figures~\ref{fig:p_kf}-\ref{fig:r_z}, we do not advocate this approach. The PCs are not a spanning set, and we do not capture the variance in the unmeasured components. This will lead us to underestimate the size of the error bars. 

In summary the adaptive grid method is a complete set for $r(z)$ and $P(k,z)$ but it is potentially sub-optimal (leading to large error bars). The PCA approach is not a complete set (leading to potential biases), but is perhaps more optimal (smaller error bars). We adopt a conservative approach not favoring the PCA approach, since in general it is better to be unbiased but have larger error bars than the other way around.

\subsection{Cosmological Inference from Adaptive Grid Reconstruction} \label{sec:cosmo agrid}

\par
\begin{figure*}
\includegraphics[width=1.\linewidth]{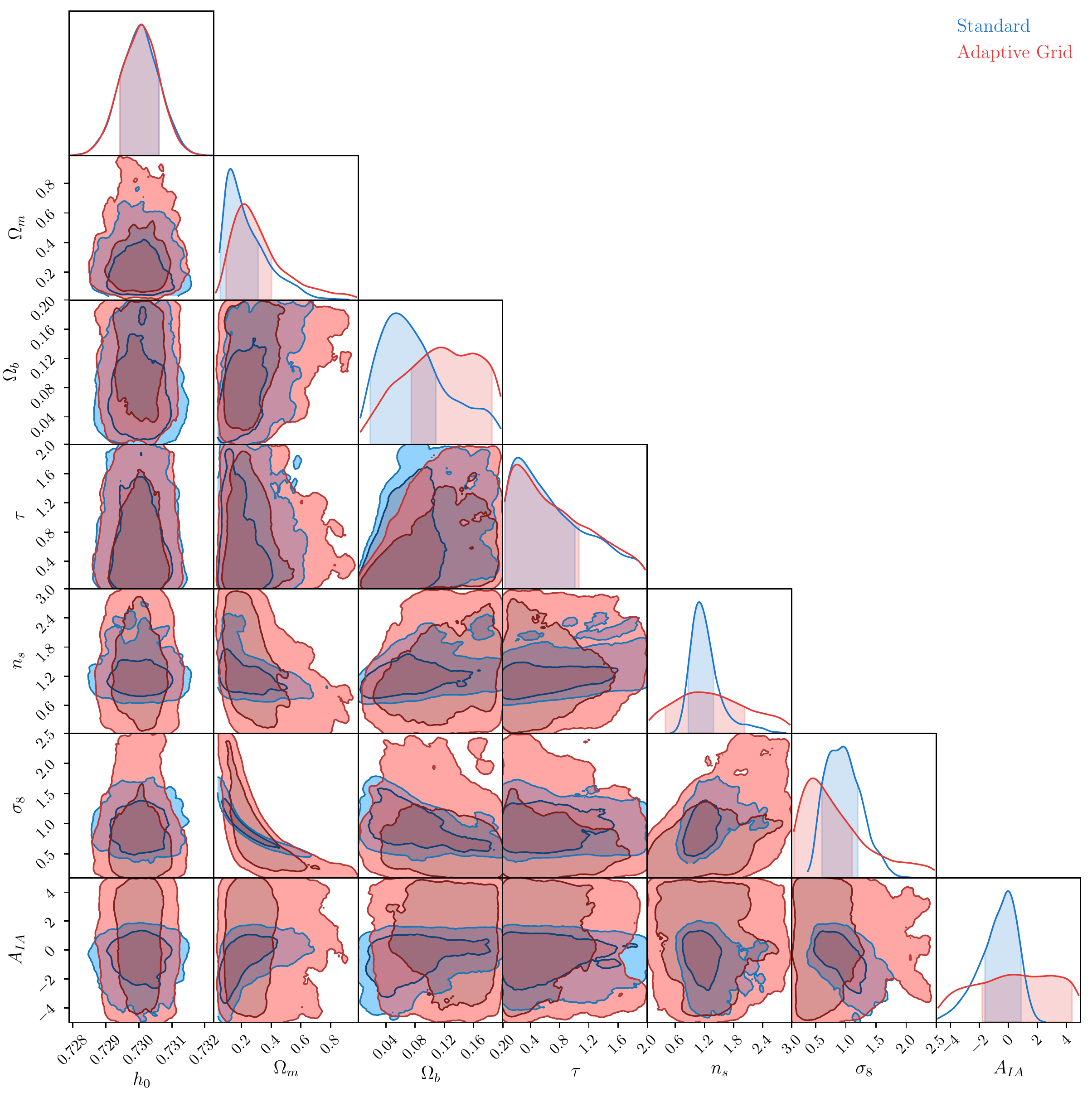}
\caption{LCDM posteriors derived from the standard cosmic shear likelihood analysis (blue) and those derived from the non-parametric information (red), using adaptive grid compression. The two techniques are in good agreement, but the posteriors in the later case are broader. This is unsurprising since information is lost in the adaptive grid compression step.}
\label{fig:grid}
\end{figure*}

 Sampling from the likelihood $\text{ln } \mathcal{L}_2 (\theta)$, we place constraints on the LCDM parameters, $\theta$, directly from the non-parametric information derived using the adaptive grid compression. We compare this to the results of the usual cosmic shear likelihood analysis by sampling from $\text{ln } \mathcal{L}_1 (\theta)$. In both cases, following the analysis of~\cite{heymans2013cfhtlens}, we place a Gaussian prior on the Hubble constant: $h_0 = 0.73 \pm 0.024$. The resulting parameter constraints are shown in Figure~\ref{fig:grid}.
\par Parameter constraints from the standard likelihood analysis are shown in blue, while constraints from the non-parametric information are in red. The two techniques are in good agreement, but the non-parametric posteriors are wider. This is expected, since information is lost in the adaptive grid compression step. Optimizing the compression step is left for a future work.

\subsection{Cosmological Inference from PCA Reconstruction}
\begin{figure*}
\includegraphics[width=1.\linewidth]{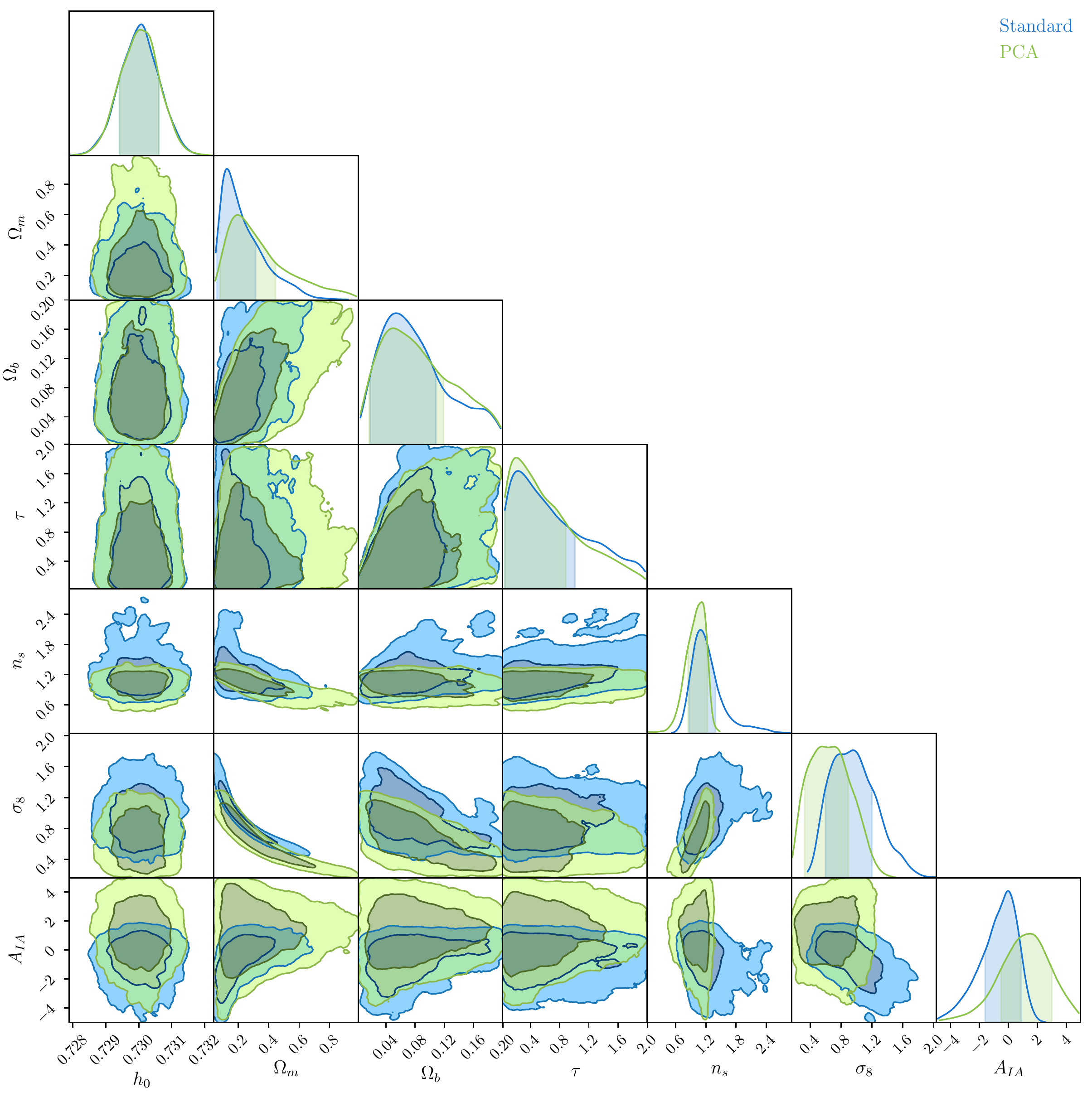}
\caption{LCDM posteriors derived from the standard cosmic shear likelihood analysis (blue) and those derived from the non-parametric information (green), using PCA compression. For the most part, the two techniques are in good agreement but there is $\sim 1 \sigma$ tension in the $\sigma_8 - \Omega_m$ plane. Since the PCs do not form a complete set, we do not capture all the variance. This leads us to underestimate error bars, and could also cause a shift in the parameter constraints. For this reason we do not advocate the PCA data compression method.}
\label{fig:pca}
\end{figure*}

By sampling from $\text{ln } \mathcal{L}_1 (\theta)$, we place constraints on the LCDM parameters directly from the non-parametric information derived using PCA compression. The resulting constraints are shown in Figure~\ref{fig:pca}, where we plot the constraints using the standard technique for comparison.
\par The constraints from the PCA compression are much tighter than those found using the adaptive grid compression, and generally agree with the posteriors from the standard cosmic shear likelihood analysis. However there is $\sim 1 \sigma$ tension in the $\sigma_8 - \Omega_m$ plane. 
\par
Discrepancies are expected in the PCA method as discussed previously. The PCs do not form a complete set, so we do not capture all the variance in the unmeasured PCs. This leads us to underestimate error bars, and could also cause a shift in the parameter constraints. For this reason we do not advocate the PCA data compression method.

\section{Future Prospects} \label{sec:future prospects}
The method we have presented has a bright future. Since the shot noise is Poissonian we expect the size of the error bars will shrink as the square root of the product of the number density and survey area. Hence the constraints on the co-moving distance and power spectrum should shrink by a factor of $\sim 2$ using Dark Energy Survey Year 1 (DESY1) data~\cite{troxel2017dark}, and by a factor of $\sim 20 - 25$ using data from a Stage IV experiment~\cite{laureijs2010euclid}.
\par In the future, comparing the non-parametric and parametric reconstructions will help us identify systematic effects and search for new physics. Discrepant power spectrum measurements might point to un-modeled baryonic physics, while a discrepant co-moving distance measurement could point to redshift dependent systematic effects such as photometric redshift errors or color dependent shear estimates -- or even a non-constant dark energy equation of state.  
\par In this reconstruction we have already identified a tension. Below $z = 0.4$, our reconstruction of the co-moving distance, $r(z)$, is larger than the Planck cosmology LCDM prediction. 
\par The next step is to repeat the analysis presented here on the DESY1 data to see if this discrepancy persists. We will fold in galaxy clustering~\cite{elvin2018dark} and galaxy-galaxy lensing~\cite{prat2018dark} into our framework. With a different set of systematic effects, these techniques will serve as a useful cross-check, as well as tightening constraints. We will also need to optimize the adaptive griding compression scheme.  
\par In the future, large experiments could repeat our analysis and release the non-parametric reconstructions as a final data product; following our technique to extract cosmological parameters from the non-parametric information would enable anyone to consistently test new physical theories without having to repeat the cosmic shear analysis or model lensing observables.
\par Constraining cosmological parameters from the non-parametric information could also be an efficient way to marginalize out small scales (high-$k$) in the matter power spectrum. These scales are difficult to model due to nonlinear growth and baryonic physics and can lead to bias. In particular, after performing the non-parametric reconstruction we can remove matter power spectrum cells at large-$k$ before using the non-parametric information to constrain the cosmological parameters. This essentially marginalizes out the small scales and this is a viable alternative, and complementary approach to, $k$-cut cosmic shear~\cite{PhysRevD.98.083514,bernardeau2014cosmic}, which explicitly removes sensitivity to small scales. We will investigate this further in the future.
\par The general philosophy in this paper was to completely separate the inference of the non-parametric information -- which is valid regardless of the underlying cosmological model -- from the inference of the cosmological parameters. This allowed us to test our non-parametric reconstruction by recovering the LCDM parameters, and a fully non-parametric reconstruction also enables the inference of parameters from any cosmological model, without having to repeat the lensing analysis. However, in the future, if the sole purpose is to search for systematics or a failure of the LCDM model, then it is more efficient to simultaneously infer the cosmological parameters $\{ \theta \}$ and a set of perturbing amplitudes $\{ \mathcal{A} \left( \theta \right) \}$. After marginalizing out the cosmological parameters, any amplitude which is not consistent with a value of one is a red flag for the presence of unidentified systematics or new physics. This simultaneous approach would allow us to get away from assuming a fiducial cosmology. 

\section{Conclusion}
We have reconstructed two lensing amplitudes, the time evolving matter power spectrum and the co-moving distance using CFHTLenS shear data. We find that the majority of the information is contained in a single lensing amplitude and the co-moving distances.
\par To reduce the dimensionality of the reconstruction problem, two different data compression regimes were employed. Although the PCA technique is efficient, the PCs do not form a spanning set, so we conclude that the adaptive grid is preferred. Optimizing the adaptive grid compression regime is left to a future work. 
\par The reconstructed co-moving distance is larger than expected from a Planck LCDM cosmology below $z = 0.4$. This could be the first sign of new physics, down to unaccounted for systematic effects or (since the tension is mild) just statistical variance. Since distance measurements from BAO and SNe Ia constrain the growth to within a few percent, unless there are large systematics in these other two probes, the last two explanations are more likely. We will investigate this further in a future work. 
\par Since a discrepancy relative to other distance measurements is only seen in the non-parametric analysis and not in the parametric one (even with a flat prior on $h_0$), comparing the non-parametric cosmic distance reconstruction measurement to BAO and SNe Ia measurements will become a powerful test to search for systematics in the shear catalog. 
\par Sampling from a Gaussian likelihood, we have shown how to extract the LCDM parameters from the non-parametric reconstruction. As well as validating the reconstruction, we expect this will be a useful method in the future. Large experiments could emulate our analysis and release  non-parametric reconstructions as a data product. This would allow theorists to consistently test new physical theories without having to repeat the lensing analysis or model lensing observables. 
\par The method presented in this work is complementary to the standard cosmic shear analysis and we advocate for its use in the analysis of upcoming cosmic shear datasets.  

\section{Acknowledgments}
\par We thank the Cosmosis team for making their code
publicly available. We thank Luke Pratley for his computing wizardry and Joe Zuntz for help with {\tt Cosmosis}. We also acknowledge useful conversations with Mark Cropper, Ignacio Ferreras and Matthew Price. We thank the anonymous referee whose comments have helped improve the paper. PT is supported by the UK Science and Technology Facilities Council. TK is supported by a Royal Society University Research Fellowship. The authors acknowledge the support of the Leverhume Trust.
\par This work is based on observations obtained with MegaPrime/MegaCam, a joint project of CFHT and CEA/IRFU, at the Canada-France-Hawaii Telescope (CFHT) which is operated by the National Research Council (NRC) of Canada, the Institut National des Sciences de l'Univers of the Centre National de la Recherche Scientifique (CNRS) of France, and the University of Hawaii. This research used the facilities of the Canadian Astronomy Data Centre operated by the National Research Council of Canada with the support of the Canadian Space Agency. CFHTLenS data processing was made possible thanks to significant computing support from the NSERC Research Tools and Instruments grant program.

\bibliographystyle{apsrev4-1.bst}
\bibliography{bibtex.bib}

\end{document}